# Some Aspects of Quantum Mechanics of Particle Motion in Static Centrally Symmetric Gravitational Fields

M.V.Gorbatenko[1], V.P.Neznamov[1,2][*], E.Yu.Popov[1]

[1]RFNC-VNIIEF, Russia, Sarov, Mira pr., 37, 607188

[2]National Research Nuclear University MEPhI, Moscow, Russia

Abstract

The domain of wave functions and effective potentials of the Dirac and Klein-Gordon equations for quantum-mechanical particles in static centrally symmetric gravitational fields are analyzed by taking into account the Hilbert causality condition.

For all the explored metrics, assuming existence of event horizons, the conditions of a "fall" of a particle to the appropriate event horizons are implemented.

The exclusion is one of the solutions for the Reissner-Nordström extreme field with the single event horizon. In this case, while fulfilling the condition found by V.I.Dokuchaev, Yu.N.Yeroshenko, the normalization integral is convergent and the wave functions become zero on the event horizon. This corresponds to the Hilbert causality condition. In our paper, due to the analysis of the effective potential for the Reissner-Nordström extreme field with real radial wave functions of the Dirac equation, the impossibility is demonstrated for the bound stationary state existence of quantum-mechanical particles, with positive energy, off the event horizon.

The analysis of the effective potential of the Dirac equation for the case of the naked singularity of the Reissner-Nordström field shown the possible existence of stationary bound states of quantum-mechanical half-spin particles.

---

[*] E-mail: neznamov@vniief.ru

## 1. Introduction

In quantum mechanics of particle motion in the external fields, the determination of the domain of the wave function $\psi(\mathbf{r},t)$ holds a prominent place.

In the current paper, when analyzing the particle motion in external centrally symmetric gravitational fields, restrictions are imposed on the domains of the appropriate wave functions following from the Hilbert causality condition [1], [2].

For the static metrics, these restrictions are validated by the analysis of the effective potentials $U(r)$ in Schrödinger relativistic one-dimensional equations derived from the appropriate systems of Dirac equations for radial functions.

The Hilbert causality condition and the analysis $U(r)$ were applied to the metrics of the Schwarzschild centrally symmetric uncharged fields in spherical coordinates $(r,\theta,\varphi)$, [3] in isotropic coordinates [4], in harmonic coordinates [5]. Historically, all the above metrics were derived due to the coordinate transformations of the Schwarzschield metrics in spherical coordinates $(r,\theta,\varphi)$ [3].

The Reissner-Nordström metric [6], [7] was also analyzed by us for the centrally symmetric charged field.

As the result of the analysis, it was found out that quantum-mechanical particles for all the examined metrics cannot cross their event horizons and are in the mode of a "fall" to the appropriate horizons.

The exception is one of the solutions for the Reissner-Nordström extreme field with the single event horizon. However, in this case as well, off the event horizon, when using the solution with real wave functions, the impossibility of existence of stationary bound states of Dirac particles with the positive energy is shown in this paper.

In sections 2, 3 of this paper, the Hilbert causality condition and the restrictions imposed by it on the wave-function domain for different metrics are discussed.

In section 4, the results of the analysis of the effective potentials $U(r)$ in Schrödinger one-dimensional relativistic equations are presented for the examined static metrics.

In the Conclusions, the results of the quantum-mechanical analysis are discussed.

In the paper, the system of units of $\hbar = c = 1$ is used; the Minkowski space signature is taken to be:

$$\eta_{\alpha\beta} = diag[1,-1,-1,-1]. \tag{1}$$

Below, as well as in (1), we will underline the indexes corresponding to the Minkowski space-time indexes.

## 2. The Hilbert Causality Condition

The causality principle was formulated by D.Hilbert in 1924 [1]:

«*So far, we have considered all the systems of coordinates $x_s$, which are derived from some system due to the arbitrary transformation, equivalent. This arbitrariness should be limited if we want to accept the point of view, according to which two world points being on the same timeline, are in the cause-and-effect relation and therefore cannot transit to synchronous world points due to the transformation. Let us identify $x_0$ as the coordinate of the **proper time** and formulate the following definition. Such a system, in which, in addition to $g < 0$, the following four inequalities are always fulfilled, is called the **eigensystem** of space-time coordinates:*

$$g_{11} < 0; \quad \begin{vmatrix} g_{11} & g_{12} \\ g_{21} & g_{22} \end{vmatrix} > 0; \quad \begin{vmatrix} g_{11} & g_{12} & g_{13} \\ g_{21} & g_{22} & g_{23} \\ g_{31} & g_{32} & g_{33} \end{vmatrix} < 0; \quad g_{00} > 0 \qquad (2)$$

and then: *"As with any transformation of coordinates, the timeline persists to be the timeline, two world points of any timeline never transit to the points with the same value of the time coordinate due to proper transformation of space-time coordinates, i.e., cannot become synchronous"*.

The causality conditions (2) are often reproduced by authors of monographs and textbooks proceeding from somewhat different considerations. For instance, in [2], the condition $g_{00} > 0$ was derived from the definition of the proper time:

$$d\tau = \sqrt{g_{00}}\,dt. \qquad (3)$$

Below, when carrying quantum-mechanical analysis and identifying the domain of wave functions, we will be governed by the fulfillment of the causality conditions (2). We will pay special attention to the fulfillment of the inequality $g_{00} > 0$; all the rest of the inequalities are fulfilled as a rule for the known solutions of General Relativity.

### 2.1 The inertial and rotating systems of coordinates in the Minkowski space.

As an example of the necessary fulfillment of the $g_{00} > 0$ condition, let us explore the Dirac Hamiltonian in the inertial and rotating frames of reference of the Minkowski space.

For the inertial frame of reference $(x'^{\mu}) = (t', x', y', z')$ the Hamiltonian of the Dirac particle, with the mass $m$ and with the unrestricted domain of wave functions, has the form of

$$H' = \gamma^0 m - i\gamma^0 \gamma^k \frac{\partial}{\partial x'^k}, \qquad (4)$$

where $\gamma^0, \gamma^k$ are four-dimensional Dirac matrices.

Let us introduce the rotating frame of reference [2]

$$t = t'; \quad x = x'\cos\omega t + y'\sin\omega t; \quad y = -x'\sin\omega t + y'\cos\omega t; \quad z = z', \qquad (5)$$

where the rate of rotation $\omega$ is the real number. The Minkowski metric in this frame of reference is stationary and has the view of

$$ds^2 = \left(1 - \omega^2(x^2 + y^2)\right)dt^2 + 2\omega(ydx - xdy)dt - dx^2 - dy^2 - dz^2. \qquad (6)$$

The Hamiltonian in the new frame of reference has the form of (see, for instance, [8])

$$H = \gamma^0 m - i\gamma^0 \gamma^k \frac{\partial}{\partial x'^k} - i\omega\left(y\frac{\partial}{\partial x} - x\frac{\partial}{\partial y}\right). \qquad (7)$$

The domain of Hamiltonian wave functions (7) is restricted by the condition $g_{00} > 0$, which for the metric (6) is reduced to the condition $\sqrt{x^2 + y^2} < \frac{1}{\omega}$. Failure to implement this condition leads to the fact that for the distance $\sqrt{x^2 + y^2} > \frac{1}{\omega}$ the rotating rate of the Dirac particle would be higher than the light velocity. So, at $g_{00} < 0$, the rotating frame of reference cannot be implemented by real bodies [2].

**3. The wave- function domains of particles with the ½ spin in centrally symmetric gravitational fields.**

**3.1 The Schwarzschild metric in the coordinates $(t, r, \theta, \varphi)$**

The square of the interval is

$$ds^2 = f_S dt^2 - \frac{dr^2}{f_S} - r^2\left(d\theta^2 + \sin^2\theta d\varphi^2\right). \qquad (8)$$

In (8), $f_S = 1 - \frac{r_0}{r}; \; r_0 = \frac{2GM}{c^2}$ is the gravitational radius (the event horizon radius), $M$ is the mass of the point spherically symmetric source of the gravitational field; $G$ is the gravitational constant; $c$ is the light velocity.

The self-conjugate Dirac Hamiltonian in $\eta$-representation with the plane scalar product of wave functions and with tetrads in the Schwinger gauge [9] for the particles with the mass $m$ and the spin ½ has the form of [10], [11]

$$H_\eta = \sqrt{f_S} m \gamma^0 - i\sqrt{f_S}\gamma^0 \left\{ \gamma^1 \sqrt{f_S}\left(\frac{\partial}{\partial r} + \frac{1}{r}\right) + \gamma^2 \frac{1}{r}\left(\frac{\partial}{\partial \theta} + \frac{1}{2}\operatorname{ctg}\theta\right) + \right.$$
$$\left. + \gamma^3 \frac{1}{r\sin\theta}\frac{\partial}{\partial \varphi} \right\} - \frac{i}{2}\frac{\partial f_S}{\partial r}\gamma^0 \gamma^1. \tag{9}$$

The domain of wave functions of the Dirac equation with the Hamiltonian (9) is restricted due to the Hilbert causality condition

$$g_{00} > 0 \to f_S = 1 - \frac{r_0}{r} > 0 \to r > r_0. \tag{10}$$

It follows from (10) that

$$\sqrt{f_S} \text{ is the positive real number.} \tag{11}$$

**3.2 Other metrics of the centrally symmetric uncharged gravitational field**

In [12], we analyze the quantum-mechanical equivalence of the metrics of a centrally symmetric uncharged gravitational field. The Schwarzschild static metrics in spherical, isotropic and harmonic coordinates, Eddington-Finkelstein [13], [14] and Painleve-Gullstrand stationary metrics [15], [16], Lemaitre-Finkelstein [17], [14] and Kruskal [18] non-stationary metrics were explored.

For each metric, the Dirac Hamiltonians were derived both directly with tetrads in the Schwinger gauge and due to coordinate transformations and the Lorentz transformations of the self-conjugate Hamiltonian in the Schwarzschild field (9).

The wave-function domains of the Dirac equation were analyzed. As the result of the analysis, it was found out that the restriction, following from the fulfillment of the Hilbert causality condition $g_{00} > 0$, on the wave-function domain of the Dirac equation in the Schwarzschild field in spherical coordinates $r > r_0$ (10) and $\sqrt{f_S} > 0$ (11), also shows itself in other coordinates for all the explored metrics:

1. the Schwarzschild metric in spherical coordinates $(t, r, \theta, \varphi)$

$$r > r_0, \tag{12}$$

2. the Schwarzschild metric in isotropic coordinates

$$R_{is} > \frac{r_0}{4}, \tag{13}$$

3. the Schwarzschild metric in harmonic coordinates

$$R_{gr} > \frac{r_0}{2}, \qquad (14)$$

4. the Eddington-Finkelstein and Painleve-Gullstrand metrics

$$r > r_0, \qquad (15)$$

5. the Lemaitre-Finkelstein metric

$$R_{L-F} - T_{L-F} > \frac{2}{3} r_0, \qquad (16)$$

6. the Kruskal metric

$$u > |v| > 0,\ u < -|v| < 0. \qquad (17)$$

It is seen from the inequalities (12) - (17) that the event horizon radius $r_0$ in the basic Schwarzschild metric (8) shows itself in new coordinates in all the explored metrics. The domains of wave functions for all the metrics do not include the singularity in the origin of coordinates.

### 3.3 The Reissner-Nordström metric

The Reissner-Nordström static metric is characterized by the point spherically symmetric sources of the gravitational field of mass $M$ and electrical field with a charge $Q$.

The square of the interval is

$$ds^2 = f_{R-N} dt^2 - \frac{dr^2}{f_{R-N}} - r^2 \left( d\theta^2 + \sin^2\theta d\varphi^2 \right), \qquad (18)$$

where $f_{R-N} = \left(1 - \frac{r_0}{r} + \frac{r_Q^2}{r^2}\right)$, $r_Q = \frac{\sqrt{G}Q}{c^2}$ - is the "charge" radius.

The Hilbert condition $g_{00} > 0$ leads to the necessity to explore positive values alone of $f_{R-N} > 0$.

1. If $r_0 > 2r_Q$, then

$$f_{R-N} = \left(1 - \frac{r_+}{r}\right)\left(1 - \frac{r_-}{r}\right), \qquad (19)$$

where $r_\pm$ is radius of inner and outer event horizons

$$r_\pm = \frac{r_0}{2} \pm \sqrt{\frac{r_0^2}{4} - r_Q^2}. \qquad (20)$$

The domain of wave functions, where $f_{R-N} > 0$, are

$$r > r_+ \ \text{and}\ r < r_-. \qquad (21)$$

2. The case $r_0 = 2r_Q$ corresponds to the Reissner-Nordström extreme field. In this case, the permitted domain is the entire space $r \in [0, \infty)$ except the radius of the single event horizon $r_\pm = \dfrac{r_0}{2}$

3. The case $r_0 < 2r_Q$ corresponds to the naked singularity. In this case, it is always $f_{R-N} > 0$ and the wave function domain is the entire space $r \in [0, \infty)$.

## 3. The analysis of effective potentials $U(\rho)$ in Schrödinger-type one-dimensional relativistic equations

The Dirac equations for fermions in centrally symmetric gravitational fields allow separation of radial and angular variables.

The spherical harmonics with the spin 1/2 are used as angular functions for centrally symmetric gravitational fields (see, for instance, [19]).

As the result, the system of equations for radial wave functions $F(\rho)$ and $G(\rho)$ in the explored external fields has the form

$$\begin{cases} \dfrac{dF(\rho)}{d\rho} = A(\rho)F(\rho) + B(\rho)G(\rho), \\ \dfrac{dG(\rho)}{d\rho} = C(\rho)F(\rho) + D(\rho)G(\rho). \end{cases} \quad (22)$$

The equations (22) are written in dimensionless variables.

$$\rho = \dfrac{r}{l_c}; \; \varepsilon = \dfrac{E}{m}; \; 2\alpha = \dfrac{r_0}{l_c} = \dfrac{2GMm}{\hbar c}; \; \alpha_Q = \dfrac{r_Q}{l_c}; \; \alpha_{em} = \dfrac{eQ}{\hbar c}; \; l_c = \dfrac{\hbar}{mc} \text{ is the Compton}$$

wavelength of the Dirac particle.

The equations (22) can be reduced to the second-order equation for the function $\psi(\rho)$, proportional either to $F(\rho)$ or to $G(\rho)$.

In the first case,

$$\psi(\rho) = F(\rho) \exp\left(\dfrac{1}{2}\int A_1(\rho')d\rho'\right). \quad (23)$$

The equation for $\psi(\rho)$ has the form of the Schrödinger equation

$$\dfrac{d^2\psi(\rho)}{d\rho^2} + 2(E_{schr} - U_{eff}(\rho))\psi(\rho) = 0. \quad (24)$$

In the equation (24),

$$E_{schr} = \frac{1}{2}(\varepsilon^2 - 1),$$

$$U_{eff}(\rho) = E_{schr} + \frac{1}{4}\frac{dA_1(\rho)}{d\rho} + \frac{1}{8}A_1^2(\rho) - \frac{1}{2}B_1(\rho).$$

(25)

In the expressions (23), (25),

$$A_1(\rho) = -\frac{1}{B(\rho)}\frac{dB(\rho)}{d\rho} - A(\rho) - D(\rho),$$

$$B_1(\rho) = -B(\rho)\frac{d}{d\rho}\left(\frac{A(\rho)}{B(\rho)}\right) - C(\rho)B(\rho) + A(\rho)D(\rho).$$

(26)

### 4.1 The effective potential in the Minkowski space with the Coulomb interaction

For the following comparison with external gravitational fields, let us consider the equation (24) with the Coulomb potential in the initial Dirac equation

$$U_D = \frac{Z\alpha_{ts}}{\rho},$$

(27)

where $Z$ is the nucleus charge, $\alpha_{ts} = \frac{e^2}{\hbar c} \approx \frac{1}{137}$ is the electromagnetic constant of the fine structure.

In this case, in the equations (22),

$$A(\rho) = -\frac{1+\kappa}{\rho}; \quad B(\rho) = \varepsilon + 1 - U_D; \quad C(\rho) = -(\varepsilon - 1 - U_D); \quad D(\rho) = -\frac{1-\kappa}{\rho},$$

(28)

where $\kappa = \pm\left(j + \frac{1}{2}\right) = \begin{cases} -(l+1), & j = l + \frac{1}{2} \\ l, & j = l - \frac{1}{2} \end{cases}$ ; $l, j$ are the quantum numbers of the orbital and the total momentum of the Dirac particle.

The asymptotic form in the origin of coordinates for the effective potential $U_C(\rho)$ has the form of

$$U_C(\rho) = -\frac{(Z\alpha_{ts})^2 - 3/4 + (1-\kappa^2)}{2\rho^2} + O\left(\frac{1}{\rho}\right).$$

(29)

The analysis of the non-relativistic quantum-mechanical problem with the potential $U(\rho) \sim -\frac{\beta}{\rho^2}$, $\beta > 0$ was carried out in [20]. Leaving out the details, let us summarize the major result: in quantum-mechanical problems no "fall" to the center occurs, if the potential at $\rho \to 0$ decrease no faster then $U(\rho) = -\dfrac{\beta}{\rho^2}$; the maximal value of $\beta$, at which no "fall" occurs yet, is $\beta_{max} = 1/8$.

Let us come back to the asymptotics (29). At $\kappa = -1$, i.e., in the s-state, the value of $\beta_{max} = 1/8$ is achieved at $Z\alpha_{ts} = 1$, i.e., at $Z \approx 137$. So, proceeding from studying the effective potential $U_C(\rho \to 0)$ alone, we arrive at the conclusion that at $Z > 137$ in relativistic one-particle problem of motion of electron in the field of the point nucleus, the quantum-mechanical mode of a "fall" to the center is implemented.

Besides, it is easy to notice that for the specified $\kappa$, such $Z_{scr}$ exist, at which the numerator in (29) changes its sign. For instance, for s, p and d states, the appropriate charges of nuclei are $Z_s = 118.7$, $Z_p = 265.3$, $Z_d = 405.4$ [†]. Figures 1, 2 present effective potentials for the s-states at the values $Z$ lower and higher than the critical value $Z_s = 118,7$.

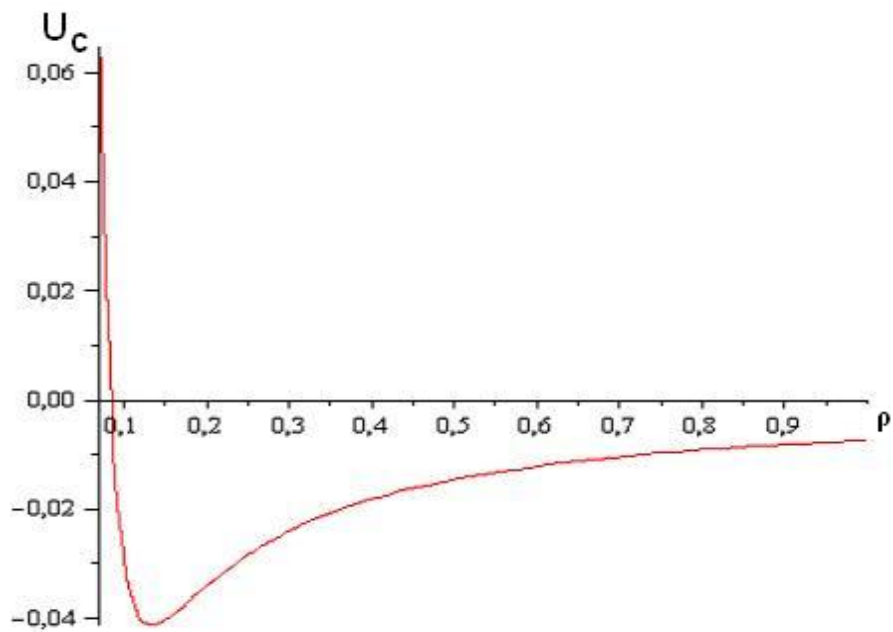

Fig.1. The effective potential $U_C(\rho)$ at the value $Z < Z_{scr}$ $\kappa = -1$, $Z\alpha_{ts} = 1/137$, $\varepsilon \approx 1$.

To the value $Z\alpha_{ts} = 1/137$, $Z \approx 1$ corresponds.

---

[†] The similar values of $Z_{scr}$ were obtained earlier from other consideration in [21], [22].

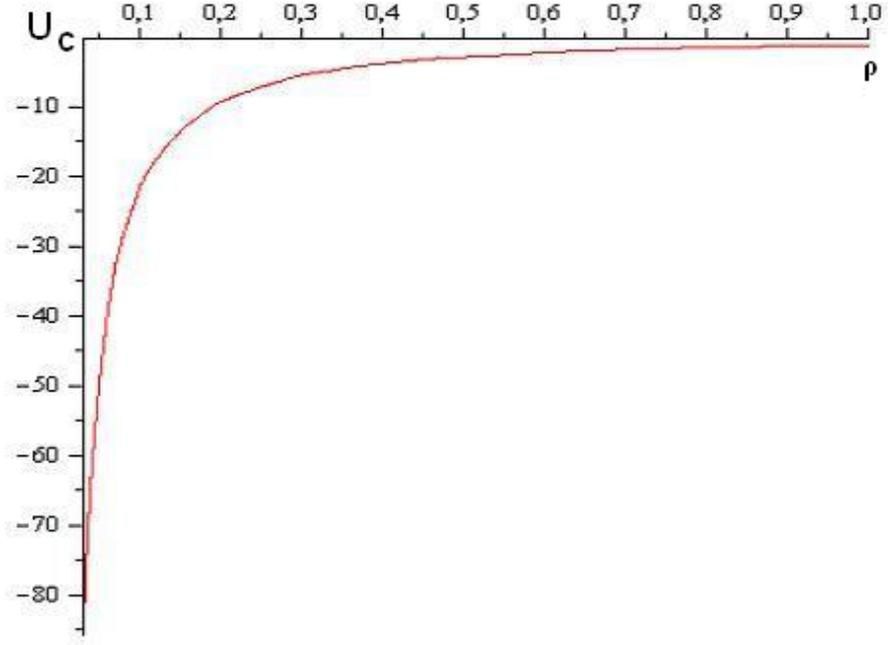

Fig.2. The effective potential $U_C(\rho)$ is at the value $Z > Z_{scr}$ $\kappa = -1$, $Z\alpha_{ts} = 0.87$, $\varepsilon = 0.49$.
To the value $Z\alpha_{ts} = 0.87$, $Z \approx 119$ corresponds.

The analysis of the expression (29) brings about the following conclusions:

- For any specified $\kappa$ there exists such a $Z_{scr}$ that at $Z<Z_{scr}$ the asymptotics of the effective potential in the origin of coordinates is positive and the potential well has a finite depth. At $Z > Z_{scr}$, the asymptotics of the effective potential changes its sign and the depth of the potential well becomes infinite, with the asymptotics having the form of $U \sim -\dfrac{\beta}{\rho^2}$; however, at $\beta < \dfrac{1}{8}$ no "fall" to the center occurs.

- For any specified $\kappa$ there is some $Z_{cr}$, starting from which value of $\beta > \dfrac{1}{8}$, that corresponds to the condition of a "fall" to the center.

Figs. 1 and 2 illustrate that at $Z < Z_{scr}$ the electron is localized in the neighborhood of minimum of the finite value potential well at an essential distance from the origin of coordinates and is separated from the point nucleus by an infinite potential barrier; at $Z \geq Z_{scr}$, firstly, the infinite potential barrier disappears, secondly, the position of the infinite-depth potential well moves by a jump to the origin of coordinates. The electron energy spectrum becomes sensitive to the behavior of the effective potential in the neighborhood of the origin of coordinates $(\rho = 0)$.

### 4.2 Effective potentials for stationary centrally symmetric uncharged gravitational fields

#### 4.2.1 The Schwarzschild field in $(r,\theta,\varphi)$ coordinates

The effective potential for the Schwarzschild field can be derived in compliance with item 4 from the system of equations for radial Dirac functions

$$\begin{cases} f_S \dfrac{dF(\rho)}{d\rho} + \left( \dfrac{1+\kappa\sqrt{f_S}}{\rho} - \dfrac{\alpha}{\rho^2} \right) F(\rho) - \left( \varepsilon + \sqrt{f_S} \right) G(\rho) = 0; \\ f_S \dfrac{dG(\rho)}{d\rho} + \left( \dfrac{1-\kappa\sqrt{f_S}}{\rho} - \dfrac{\alpha}{\rho^2} \right) G(\rho) + \left( \varepsilon - \sqrt{f_S} \right) F(\rho) = 0. \end{cases} \quad (30)$$

The leading singularity of the effective potential in the vicinity of the event horizon has the form of

$$U_S(\rho \to 2\alpha) = -\frac{\frac{1}{8} + 2\alpha^2 \varepsilon^2}{(\rho - 2\alpha)^2}. \quad (31)$$

The numerator of the equation (31) is higher than 1/8 and in compliance with item 4.1 any quantum-mechanical particle will move in the Schwarzschild field in the mode of a "fall" to the horizon $\rho = 2\alpha$ at any values of $\alpha$ and $\varepsilon$.

#### 4.2.2 The Schwarzschild field in isotropic coordinates

The coordinates are

$$(t, R, \theta, \varphi). \quad (32)$$

The square of the interval is

$$ds^2 = V^2(R) dt^2 - W^2(R) \left[ dR^2 + R^2 \left( d\theta^2 + \sin^2\theta d\varphi^2 \right) \right]. \quad (33)$$

Here,

$$V(R) = \frac{1 - \dfrac{r_0}{4R}}{1 + \dfrac{r_0}{4R}}, \quad W(R) = \left( 1 + \dfrac{r_0}{4R} \right)^2, \quad F_{Ob} = \frac{V(R)}{W(R)} = \frac{1 - \dfrac{r_0}{4R}}{\left( 1 + \dfrac{r_0}{4R} \right)^3}. \quad (34)$$

The domain of variable $R$ is

$$R > \frac{r_0}{4}. \quad (35)$$

The self-conjugate Dirac Hamiltonian has the form of [23], [12]

$$H_\eta = V(R) \beta m + \frac{1}{2} \left( \boldsymbol{\alpha} \mathbf{p} F_{Ob}(R) + F_{Ob}(R) \boldsymbol{\alpha} \mathbf{p} \right). \quad (36)$$

In (36), $\alpha^k, \beta$ are Dirac matrixes, $p^k = -i\dfrac{\partial}{\partial R^k}$ - are the momentum operators of the Dirac particle.

After separation of the variables, the system of equations for radial functions $F_{is}(\rho)$ and $G_{is}(\rho)$ has the view of

$$F_{Ob}\frac{dF_{is}(\rho)}{d\rho} + \left(F_{Ob}\frac{1+\kappa}{\rho} + \frac{1}{2}\frac{dF_{Ob}}{d\rho}\right)F_{is}(\rho) - (\varepsilon + V(\rho))G_{is}(\rho) = 0,$$

$$F_{Ob}\frac{dG_{is}(\rho)}{d\rho} + \left(F_{Ob}\frac{1-\kappa}{\rho} + \frac{1}{2}\frac{dF_{Ob}}{d\rho}\right)G_{is}(\rho) + (\varepsilon - V(\rho))F_{is}(\rho) = 0. \quad (37)$$

The leading singularity of the effective potential in the vicinity of the event horizon $\dfrac{r_0}{4l_c} = \dfrac{\alpha}{2}$ has the view of

$$U_{is}\left(\rho \to \frac{\alpha}{2}\right) = -\frac{\frac{1}{8} + 8\alpha^2\varepsilon^2}{\left(\rho - \dfrac{\alpha}{2}\right)^2}. \quad (38)$$

The numerator (38) is higher than $\frac{1}{8}$ at any values of $\alpha$ and $\varepsilon$.

**4.2.3 The Schwarzschild field in spherical harmonic coordinates**

The coordinates are

$$(t, R, \theta, \varphi). \quad (39)$$

The interval square is

$$ds^2 = \frac{\left(R - \dfrac{r_0}{2}\right)}{\left(R + \dfrac{r_0}{2}\right)}dt^2 - \frac{\left(R + \dfrac{r_0}{2}\right)}{\left(R - \dfrac{r_0}{2}\right)}dR^2 - \left(R + \frac{r_0}{2}\right)^2(d\theta^2 + \sin^2\theta d\varphi^2). \quad (40)$$

The domain of variable $R$ is

$$R > \frac{r_0}{2}. \quad (41)$$

The self-conjugate Hamiltonian of the particles with the spin ½ has the form of [12]

$$H_\eta = \sqrt{F_g}\beta m - i\alpha^1\left[F_g\left(\frac{\partial}{\partial R} + \frac{1}{R}\right) + \frac{r_0}{2R^2\left(1 + \dfrac{r_0}{2R}\right)^2}\right] -$$

$$-i\sqrt{F_g}\frac{1}{1 + \dfrac{r_0}{2R}}\left[\alpha^2\frac{1}{R}\left(\frac{\partial}{\partial \theta} + \frac{1}{2}\operatorname{ctg}\theta\right) + \alpha^3\frac{1}{R\sin\theta}\frac{\partial}{\partial \varphi}\right]. \quad (42)$$

In (42),

$$F_g(r) = \frac{1 - \frac{r_0}{2R}}{1 + \frac{r_0}{2R}}. \tag{43}$$

The system of equations for the radial functions $F_{gr}(\rho), G_{gr}(\rho)$ has the view of

$$F_g \frac{dF_{gr}(\rho)}{d\rho} + \left( F_g \frac{1}{\rho} + \frac{\sqrt{F_g}}{\left(1 + \frac{\alpha}{\rho}\right)} \frac{\kappa}{\rho} + \frac{1}{2} \frac{dF_g}{d\rho} \right) F_{gr}(\rho) - \left(\varepsilon + \sqrt{F_g}\right) G_{gr}(\rho) = 0,$$

$$F_g \frac{dG_{gr}(\rho)}{d\rho} + \left( F_g \frac{1}{\rho} - \frac{\sqrt{F_g}}{\left(1 + \frac{\alpha}{\rho}\right)} \frac{\kappa}{\rho} + \frac{1}{2} \frac{dF_g}{d\rho} \right) G_{gr}(\rho) + \left(\varepsilon - \sqrt{F_g}\right) F_{gr}(\rho) = 0. \tag{44}$$

The leading singularity of the effective potential in the vicinity of the event horizon $\frac{r_0}{2l_c} = \alpha$ has the view of

$$U_{gr}(\rho \to \alpha) = -\frac{\frac{1}{8} + 2\alpha^2 \varepsilon^2}{(\rho - \alpha)^2}. \tag{45}$$

The numerators of the expressions (38) and (45) is higher than 1/8 at any value of $\alpha$ and $\varepsilon$.

Any quantum-mechanical particle with the spin 1/2 in the Schwarzschild fields in isotropic and harmonic coordinates as well as in the Schwarzschild field in spherical coordinates will move in the mode of a "fall" to the appropriate event horizon.

The quantum mechanics of motion of zero-spin particles must follow the same laws. Let us illustrate it with the example of motion of the zero-spin particle with the mass $m$ in the Schwarzschild field.

**4.3 The effective potential of the Klein-Gordon equation for the Schwarzschild field in the spherical coordinates $(r, \theta, \varphi)$**

The Klein-Gordon equation for the Schwarzschild metric has the form of

$$(-g)^{1/2} \frac{\partial}{\partial x^\mu} \left[ (-g)^{1/2} g^{\mu\nu} \frac{\partial}{\partial x^\nu} \Phi \right] + m^2 \Phi = 0. \tag{46}$$

In (46) $(-g)^{1/2} = r^2 \sin\theta$.

Upon separation of the variables by using the substitution

$$\Phi(r, \theta, \varphi, t) = R(r) Y(\theta, \varphi) e^{-i\omega t}, \tag{47}$$

where $Y(\theta, \varphi)$ are spherical harmonics, we obtain the equation for the radial wave

functions $R(r)$

$$\frac{d^2R}{dr^2} + A(r)\frac{dR}{dr} + B(r)R = 0, \qquad (48)$$

where $A(r) = \frac{2r-r_0}{r^2}\frac{1}{f_s}$, $B(r) = -\frac{1}{f_s r^2}\left(l(l+1) - \frac{\omega^2 r^2}{f_s} + m^2 r^2\right)$, $f_s = 1 - \frac{r_0}{r}$, $l$ is the quantum number of the orbital moment of the scalar particle.

After substitution of $R(r) = Z \cdot U$, where $U(r) = \exp\left(-\frac{1}{2}\int A(r')dr'\right)$, we will obtain the Schrödinger-type equation with the effective potential $U(r)$

$$\frac{d^2 Z}{dr^2} + \kappa^2(r) Z = 0, \qquad (49)$$

where $\kappa^2(r) = 2(E_{Schr} - U(r)) = -\frac{1}{2}\frac{dA}{dr} - \frac{1}{4}A^2 + B$.

As the result, the asymptotics of the effective potential in the vicinity of the event horizon coincides with the similar asymptotics for the Dirac particle (31) and has the view of

$$U(r \to r_0) = -\frac{\frac{1}{8} + 2\left(\frac{r_0}{2}\right)^2 \omega^2}{(r-r_0)^2};$$

in dimensionless variables:

$$U(\rho \to 2\alpha) = -\frac{\frac{1}{8} + 2\alpha^2 \omega^2}{(\rho - 2\alpha)^2}. \qquad (50)$$

**4.4 The effective potentials for the Reissner-Nordström centrally symmetric charged field**

The Reissner-Nordström metric is presented in the expression (18).

The self-conjugate Hamiltonian of half-spin particle with mass $m$ and charge $e$ has the form of [24]

$$H_\eta = \sqrt{f_{R-N}}\beta m - i\alpha^1\left(f_{R-N}\frac{\partial}{\partial r} + \frac{1}{r} - \frac{r_0}{2r^2}\right) - \\ -i\sqrt{f_{R-N}}\frac{1}{r}\left[\alpha^2\left(\frac{\partial}{\partial \theta} + \frac{1}{2}\text{ctg}\theta\right) + \alpha^3 \frac{1}{\sin\theta}\frac{\partial}{\partial \varphi}\right] + \frac{eQ}{r}. \qquad (51)$$

In (51) $\alpha^k, \beta$ are Dirac metrics.

After separation of the variables, the system of equations for radial functions $F_{R-N}(\rho), G_{R-N}(\rho)$ has the view of

$$f_{R-N}\frac{dF_{R-N}(\rho)}{d\rho}+\left(\frac{1+\kappa\sqrt{f_{R-N}}}{\rho}-\frac{\alpha}{\rho^2}\right)F_{R-N}(\rho)-\left(\varepsilon-\frac{\alpha_{em}}{\rho}+\sqrt{f_{R-N}}\right)G_{R-N}(\rho)=0,$$
$$f_{R-N}\frac{dG_{R-N}(\rho)}{d\rho}+\left(\frac{1-\kappa\sqrt{f_{R-N}}}{\rho}-\frac{\alpha}{\rho^2}\right)G_{R-N}(\rho)+\left(\varepsilon-\frac{\alpha_{em}}{\rho}-\sqrt{f_{R-N}}\right)F_{R-N}(\rho)=0.$$
(52)

In (52), the dimensionless variable has been introduced:

$$\rho=\frac{r}{l_c};\ \varepsilon=\frac{E}{m};\ 2\alpha=\frac{r_0}{l_c}=\frac{2GMm}{\hbar c};\ \alpha_Q=\frac{r_Q}{l_c}=\frac{\sqrt{G}QM}{\hbar c};\ \alpha_{em}=\frac{eQ}{\hbar c};\ l_c=\frac{\hbar}{mc}\ \text{is the Compton}$$

wave length of the Dirac particle.

The value $f_{R-N}$ can be presented in the form of

$$f_{R-N}=1-\frac{2\alpha}{\rho}+\frac{\alpha_Q^2}{\rho^2}=\left(1-\frac{\rho_+}{\rho}\right)\left(1-\frac{\rho_-}{\rho}\right),\quad(53)$$

where

$$\rho_+=\alpha+\sqrt{\alpha^2-\alpha_Q^2}\ -\ \text{is the outer event horizon radius}\quad(54)$$

$$\rho_-=\alpha-\sqrt{\alpha^2-\alpha_Q^2}\ -\ \text{is the inner event horizon radius.}\quad(55)$$

Let us explore three cases of relationship between $\alpha$ and $\alpha_Q$.

**4.4.1 $\alpha^2>\alpha_Q^2$**

In this case, the wave function domain of Hamiltonian (51) has the view of

$$\rho\in[0,\rho_-)\ \text{and}\ \rho\in(\rho_+,\infty).\quad(56)$$

The effective potentials in this case have the following leading singularities at $\rho\to\rho_-\ (\rho<\rho_-)$ and $\rho\to\rho_+\ (\rho>\rho_+)$

$$U_{R-N}(\rho\to\rho_-)=-\frac{1}{8(\rho_--\rho)^2}\left[1+\frac{(\varepsilon\rho_--\alpha_{em})^2\rho_-^2}{(\alpha-\rho_-)^2}\right],\quad(57)$$

$$U_{R-N}(\rho\to\rho_+)=-\frac{1}{8(\rho_+-\rho)^2}\left[1+\frac{(\varepsilon\rho_+-\alpha_{em})^2\rho_+^2}{(\rho_+-\alpha)^2}\right].\quad(58)$$

The numerators in the expressions (57), (58) are always $\geq\frac{1}{8}$ at any value of $\alpha,\alpha_Q,\alpha_{em}$ and $\varepsilon$. If the particle motion is explored in the domain $\rho\in[0,\rho_-)$, such a particle will be in the mode of a "fall" to the inner event horizon. When exploring the motion of the particles in the domain $\rho\in(\rho_+,\infty)$, the quantum-mechanical particle $\alpha$ will be in the mode of a "fall" to the outer event horizon.

**4.4.2 $\alpha^2 = \alpha_Q^2$ is the Reissner-Nordström extreme field**

In this case, there is the single event horizon with radius $\rho_+ = \rho_- = \alpha$.

The domain of wave functions is the entire space $\rho \in [0, \infty)$, except for the event horizon radius $\rho_\pm = \alpha$, on which $g_{00} = 0$.

If $\varepsilon \neq \dfrac{\alpha_{em}}{\alpha}$, the effective potential has the leading singularity at $\rho \to \alpha$.

$$U_{R-N}^{extr}(\rho \to \alpha) = -\dfrac{\left(\varepsilon - \dfrac{\alpha_{em}}{\alpha}\right)^2 \alpha^4}{8(\rho - \alpha)^4}. \tag{59}$$

For $U(\rho \to \alpha) \sim \dfrac{1}{(\rho - \alpha)^4}$ the quantum-mechanical particle, being above or below the event horizon, will move in the mode of a "fall" to the horizon $\rho_\pm = \alpha$ [20].

At $\varepsilon = \dfrac{\alpha_{em}}{\alpha}$, the effective potential has the view of

$$\left(U_{R-N}^{extr}\right)_1(\rho) = \dfrac{1}{2}\left[\dfrac{\left(1 - \dfrac{\alpha_{em}^2}{\alpha^2}\right)\rho^4 + (\kappa^2 + \kappa)\rho^2 - \alpha(\kappa+1)\rho + \dfrac{3}{4}\alpha^2}{\rho^2(\rho - \alpha)^2} - \left(1 - \dfrac{\alpha_{em}^2}{\alpha^2}\right)\right]. \tag{60}$$

The leading singularity of the potential (60) becomes

$$\left(U_{R-N}^{extr}\right)_1(\rho \to \alpha) = -\dfrac{\frac{1}{4} - \kappa^2 - \alpha^2 + \alpha_{em}^2}{2(\rho - \alpha)^2}. \tag{61}$$

For the existence of the solution $\varepsilon = \dfrac{\alpha_{em}}{\alpha}$ with the converging normalization integral for wave functions, the fulfillment of the following condition is needed

$$\kappa^2 + \alpha^2 - \alpha_{em}^2 > \dfrac{1}{4}. \tag{62}$$

The solution $\varepsilon = \dfrac{\alpha_{em}}{\alpha}$ with the condition (62) was derived earlier in [25] due to the direct solution for the system of equations for radial functions of the Dirac equation.

From the solution $\varepsilon = \dfrac{\alpha_{em}}{\alpha}$ it follows that the positive energy values of the Dirac particle are implemented at the same signs of charges $Q$ and $e$ and, vice-versa, the negative energy values $\varepsilon$ are implemented at opposite signs of the charges $Q$ and $e$.

Confining oneself to the solutions with positive energies off the event horizon, the value $\varepsilon = \dfrac{\alpha_{em}}{\alpha}$ will be the solution for the bound states of the Dirac particle, if $\varepsilon < 1$, i.e.,

$$\alpha_{em} < \alpha. \tag{63}$$

For the domain of wave functions $\rho \in (\alpha, \infty)$, at $\alpha_{em} < \alpha$, the effective potential (60) will be positive everywhere and according to the results of the numerical calculations contains no extremums. The asymptotics of the potential (60) at $\rho \to \infty$

$$\left(U_{R-N}^{extr}\right)(\rho \to \infty) = 0. \tag{64}$$

There is no potential well for the particle and hence there are no stationary bound states for the Dirac particle off the event horizon $\rho_\pm = \alpha$. The similar conclusion was drawn earlier in [26].

The effective potential (60) under the event horizon $\rho \in [0, \alpha)$ is positive everywhere with the asymptotics $\sim \left.\dfrac{1}{\rho^2}\right|_{\rho \to 0}$ and $\sim \left.\dfrac{1}{(\rho-\alpha)^2}\right|_{\rho \to \alpha}$. The energy level for the bound states $\varepsilon = \dfrac{\alpha_{em}}{\alpha}$ has to be also positive. It is achieved at the same charge signs of $Q$, $e$ and at

$$\dfrac{\alpha_{em}}{\alpha} > \min\left(U_{R-N}^{extr}\right)_1(\rho).$$

### 4.4.3 $\alpha^2 < \alpha_Q^2$ is the naked singularity

In this case, the event horizons disappear, the value $f_{R-N} = 1 - \dfrac{2\alpha}{\rho} + \dfrac{\alpha_Q^2}{\rho^2}$ is always higher than zero.

The wave-function domain is the entire space of $\rho \in [0, \infty)$.

The effective potential at $\rho \to 0$ is positive and has the view of

$$\left(U_{R-N}^{n-s}\right)(\rho \to 0) = \dfrac{3}{8}\dfrac{1}{r^2}. \tag{65}$$

The full form of $\left(U_{R-N}^{n-s}\right)(\rho)$ at some values of $\alpha, \alpha_Q, \alpha_{em}$ qualitatively coincides with the form of the effective potential of the Coulomb field $U_c(\rho)$ at $Z \leq 119$ (see fig. 1). It testifies to the possible existence of stationary bound states of quantum-mechanical half-spin particles for naked singularity of the Reissner-Nordström field.

## 5. Conclusions

The analysis of the effective potentials of the Dirac equation in static centrally symmetric gravitational fields allows the following conclusions:

1. For all the explored metrics allowing event horizons, the motion of quantum-mechanical particles is implemented in the mode of a "fall" to the appropriate event horizons. In this mode, the particle cannot cross the event horizons of the explored metrics. This is in full agreement with the restrictions on the wave function domains imposed by the Hilbert causality condition.

2. The absence of the mode of a "fall" of a particle to the event horizon is implemented in the only case of the Reissner-Nordström extreme field $\left(\alpha = \alpha_Q;\ \rho_+ = \rho_- = \alpha\right)$ for the solution $\varepsilon = \dfrac{\alpha_{em}}{\alpha}$, found by the authors [25]. In this case, with fulfillment of the condition (62) the normalization integral is convergent and the wave functions of the Dirac equation turn to zero on the single event horizon $\rho_+ = \rho_- = \alpha$. The above factor corresponds to the Hilbert causality condition. For the bound states of particles, with positive energy (i.e. at the same sings of charges $Q, e$), off the event horizon, the condition of $\alpha^{em} < \alpha$ must be met. In this case, the behavior of the effective potential (60), off the event horizon, leads to the absence of existence of stationary bound states of Dirac particles.

3. For the case of the naked singularity of the Reissner-Nordström field $\left(\alpha^2 < \alpha_Q^2\right)$, the outer and inner horizons of events are absent, the form of the effective potential (65) shows the possible existence of stationary bound states of quantum-mechanical half-spin particles.

In the available monographs and textbooks of relativistic quantum mechanics when the conditions of a "fall" of a quantum-mechanical-particle to the center are fulfilled, the impossibility is stated for the following exploration of their behavior within the framework of the single-particle theory due to the spontaneous production of pairs "particle-antiparticle" taking place in these condition (see, for instance, [27]). However, in [21], [22], it is shown, with regard to motion of electron in the Coulomb field of heavy nuclei with $Z > 137$, that in this case as well, when the mode of a "fall" of a particle to the center is implemented, one can correctly formulate, from the viewpoint of mathematics, the solution of quantum-mechanical problems within the framework of the single-particle relativistic quantum theory. As the result, in [21], [22] for the Coulomb potentials with any value of $Z$, parametric families of relativistic self-conjugate Hamiltonians of Dirac equations were found. In order to single out the only Hamiltonian from these families, an additional physical condition is needed. Such a condition for this problem is the accounting of finite dimensions of the nuclei [28], [29], [30]. The approach of [21], [22] can be applied to the quantum-mechanics exploration of particle motion in centrally symmetric gravitational fields.

**Acknowledgements**

The authors would like to thank Prof. P.Fiziev for the useful discussions and A.L. Novoselova for the substantial technical help in the preparation of this paper.